# Diode-pumped Pr:BaY$_2$F$_8$ cw orange laser

**David Pabœuf,**[1,2] **Oussama Mhibik,**[1] **Fabien Bretenaker,**[1,*] **Philippe Goldner,**[2] **Daniela Parisi**[3] **and Mauro Tonelli**[3]

[1]*Laboratoire Aimé Cotton, CNRS-Université Paris Sud 11, 91405 Orsay Cedex, France*

[2]*Laboratoire de Chimie de la Matière Condensée de Paris, CNRS-UMR 7574, Chimie Paristech 11 rue Pierre et Marie Curie, 75005 Paris, France*

[3]*NEST, Nanoscience Institute –CNR, Dipartimento di Fisica, Università di Pisa, Largo B. Pontecorvo,3 -56127- Pisa, Italy*

[*]*Corresponding author: fabien.bretenaker@lac.u-psud.fr*

We report the realization of the continuous wave laser emission in the orange at 607 nm from a Pr:BaY$_2$F$_8$ (Pr:BYF) crystal pumped by a blue GaN laser diode. A maximal output power of 78 mW is obtained in a quasi single transverse mode beam. The effect of reabsorption losses at the laser wavelength is also evidenced.

Pr$^{3+}$-doped materials offer a large range of available laser transitions in the visible spectrum ranging from blue to red. For a long time, the main difficulty with Pr-doped lasers was the pumping scheme. Indeed, the absorption transitions of Pr$^{3+}$ lie in the blue between 440 and 480 nm and the first demonstrations of such a laser have been made by pumping either with an Ar$^+$ laser [1,2], a dye laser [3] or, more recently, a frequency doubled optically pumped semiconductor laser (OPSL) [4]. The recent developments of high power blue laser diodes emitting around 445 nm enable nowadays to revisit these Pr-doped materials and build efficient Pr-doped lasers in compact setups. Triggered by the applications for display, most of the laser transitions that were investigated in the literature concern the green emission around 520 nm and the red emission around 640 nm in various fluoride matrices [5-7]. Laser emission in the deep red at 747 nm has also recently been demonstrated with Pr:YAlO$_3$ [8].

However, the orange laser emission on the $^3P_0 \rightarrow {}^3H_6$ transition of Pr$^{3+}$ around 600 nm may be of a particular interest since there is a lack of laser sources in this wavelength range. Indeed, there is no laser diode emitting at these wavelengths and the main available sources are dye lasers and optical parametric oscillators [9]. An orange diode-pumped Pr-doped laser would then offer a compact and efficient all-solid-state alternative to these heavy systems. Only a few works have been reported on orange Pr$^{3+}$ lasers with Pr:LiYF$_4$ and Pr:LiLuF$_4$ [10, 11], the pumping source being in both cases a frequency doubled OPSL. Orange emission has also been obtained from a Pr:BaY$_2$F$_8$ (Pr:BYF) crystal co-doped with Yb$^{3+}$ in an avalanche up-conversion process under Ti:Sapphire pumping around 820 and 840 nm [12,13]. In this letter, we present the realization of a diode pumped Pr:BYF laser emitting at 607 nm which is, to the best of our knowledge, the first diode pumped Pr-doped laser emitting in the orange.

Our experimental setup is described in Figure 1. The pump laser source consists of a commercially available spatially multimode GaN laser diode. At maximum current, it provides 1W with an emission centered at 444 nm and a spectral total width at -10dB of 2 nm. This beam is linearly polarized in the direction parallel to the junction. After collimation by a high numerical aperture aspherical lens (Geltech 352610, focal length : $f_1$ = 4 mm, N.A. = 0.6), the beam is elliptical with an aspect ratio of 1/4. In addition, this beam is not perfectly collimated and presents a slight astigmatism: a pair of cylindrical lenses (focal lengths : $f_{cyl1}$ = -20 mm and $f_{cyl2}$ = 50 mm) is then necessary to correct it and collimate the beam in both directions, the focal lengths being chosen to optimize the circularization.

We first checked the beam quality of the pump by observing its intensity profile with a CCD camera (Spiricon SP-980M) positioned after the first collimating lens (see Figure 2). This picture evidences the poor beam quality, especially in the fast axis direction. Indeed, in the slow axis direction, the few lobes that can be seen have to be expected since the laser diode waveguide is multimode in that direction. In contrast, in the fast axis direction, such a laser diode is supposed to be single mode, which is clearly not the case in view of the 6 horizontal ripples we observe.

The laser cavity is a quasi-concentric concave-concave cavity composed of two meniscus mirrors with a radius of curvature of 50 mm. The input mirror ($M_1$) is highly transmittive for the pump wavelength (T>95% @ 444 nm) and highly reflective at the laser line (R>99.9% @ 607 nm). Three different transmittances have been used for the output coupler ($M_2$): T = 0.1%, 1% and 3.5%. Moreover, special care has been taken to avoid laser oscillation from the $^3P_0 \rightarrow {}^3F_2$ red transition around 640 nm by specifying a sufficiently high transmittance in the red on both cavity mirrors. We used a 5.67 mm-long crystal with a $Pr^{3+}$ doping concentration of 1.25% (in the melt). This crystal was cut along the optical axes, X, Y and Z and mounted on a simple copper plate with no specific cooling. The crystal was oriented with the Y-axis parallel to the polarization of the pump excitation in order to maximize pump absorption [14]. Antireflection coatings (R<0.1%) for both pump and laser wavelengths were deposited on its facets.

Different pump beam sizes inside the Pr:BYF crystal have been tested to optimize the laser performance. This was done by choosing proper focal lengths for lenses $L_1$ and $L_2$ to adapt the magnification of the telescope as well as for the focusing lens $L_4$ (see Figure 1). The best results have been obtained with the following focal lengths: $f_2 = 100$ mm, $f_3 = 300$ mm and $f_4 = 75$ mm, resulting in an almost circular pump beam with a mean diameter of 70 μm inside the crystal.

In that pumping configuration, the small signal absorption of the crystal has been measured to be 73%. Orange laser oscillation has been obtained in continuous regime. The laser spectrum has been measured with an optical spectrum analyzer with a resolution of 10 pm. The central wavelength was found to be 607.2 nm with a total spectral width at -3dB of 25 pm. The beam is linearly polarized parallel to the Y-axis direction.

Figure 3 shows the output power characteristics with different transmittances of the output coupler. From the laser thresholds and optical efficiencies obtained with the three different output coupling values, the round trip losses in the resonator have been evaluated by use of the Findlay-Clay [15] and Caird [16] analyses, which led respectively to values of 3.4% and 3.2% for the total internal round trip losses inside the cavity (not including the transmission of the output coupler). Such high losses are common facts for orange emitting Pr lasers [10,11]. The explanation that was proposed is reabsorption at the laser wavelength on a transition from one of the Stark levels of the ground state to the $^1D_2$ level (see for example ref [17] for a detailed analysis of the energy levels of Pr:BYF). Such an absorption at the laser wavelength would indeed limit the laser efficiency in the same way as it affects quasi-3-level lasers. In order to confirm this effect and evaluate its incidence on the laser process, we have measured the single-pass absorption with another Pr:BYF sample and our orange laser beam. The Pr:BYF sample was 4.85 mm long and oriented in the same way as the laser crystal inside the cavity. We measured a single-pass absorption of 1.4 %. This would correspond to a double-pass absorption of 3.2 % for the 5.67 mm-long crystal we used in the laser cavity, which is very close to the values deduced from the Findlay-Clay and Caird analyses.

The 3.5% transmitting output coupler led to the best results with a laser output power of 78 mW and a laser slope efficiency with respect to the incident pump power of 13 %. From the previous losses analysis, we can roughly estimate that the optimum output coupling value in our case would be 5 % which would lead to a laser output power around 100 mW for 1 W of pump power. The modest value of the laser slope efficiency is mostly attributed to the poor beam quality of the pump beam which reduces the overlap between the laser and the pump beams. A significant increase of both slope efficiency and output power should then be expected with an improved pump laser diode.

In order to characterize the beam quality of the orange laser, we have measured its $M^2$ factor by measuring the evolution of the laser beam after imaging by a f=100 mm lens using the knife-edge technique. The resulting measurements are shown on Figure 4. The fit of these data leads to a $M^2$ value of 1.2. A measurement of the beam profile intensity is shown in the inset in Figure 4. It clearly evidences the good beam quality and its quasi-Gaussian shape.

In conclusion, we have reported the realization of a diode pumped orange emitting Pr:BaY$_2$F$_8$ laser. A maximum output power of 78 mW has been obtained at 607 nm in a quasi-single transverse mode. The limiting effect of losses induced by reabsorption at the laser wavelength on the $^3H_4 \rightarrow {}^1D_2$ transition of $Pr^{3+}$ has also been evidenced. Investigations on the incidence of the temperature on the reabsorption are currently under progress to increase the laser efficiency.

This first realization of a diode-pumped Pr-doped laser emitting in the orange opens the possibility to build compact and efficient laser sources and fill the gap in this spectrum range where no laser diode is available. This may also open new applications for diode-pumped Pr-doped laser such as quantum processing of $Pr^{3+}$ in $Y_2SiO_5$ for which a narrow-linewidth frequency-stabilized laser emitting at 606 nm is required [18,19]. Among Pr-doped fluoride crystals, we have identified Pr:KYF$_4$ as a promising candidate to reach this wavelength. This will be the concern of our future investigations.

This work has been partially funded through the ANR project FLUOLASE and by the Triangle de la Physique. The authors are thankful to P. Camy for fruitful discussions and to C. Siour for his technical support with the experiments.

17. B.E. Bowlby, B. Di Bartolo, Journal of Luminescence **100**, 131-139 (2002)

18. L. Rippe, M. Nilsson, S. Kröll, R. Klieber, and D. Suter, Physical Review A **71**, 1-12 (2005).

19. M. P. Hedges, J. J. Longdell, Y. Li, and M. J. Sellars, Nature **465**, 1052-1056 (2010).

**References with title:**

1. R. Smart, J. Carter, A. Tropper, D. Hanna, S. Davey, S. Carter, and D. Szebesta, "CW room temperature operation of praseodymium-doped fluorozirconate glass fibre lasers in the blue-green, green and red spectral regions, "Optics Communications **86**, 333-340 (1991).

2. T. Sandrock, T. Danger, E. Heumann, G. Huber, and B. H. Chai, "Efficient Continuous Wave-laser emission of $Pr^{3+}$-doped fluorides at room temperature, "Applied Physics B **58**, 149-151 (1994).

3. T. Danger, T. Sandrock, E. Heumann, G. Huber, and B. H. Chai, "Pulsed laser action of $Pr:GdLiF_4$ at room temperature, "Applied Physics B Photophysics and Laser Chemistry **57**, 239-241 (1993).

4. A. Richter, N. Pavel, E. Heumann, G. Huber, D. Parisi, A. Toncelli, M. Tonelli, A. Diening, and W. Seelert, "Continuous-wave ultraviolet generation at 320 nm by intracavity frequency doubling of red-emitting Praseodymium lasers, "Optics Express **14**, 3282-3287 (2006).

**Figure captions list :**

Figure 1 : Experimental set-up. DL : diode laser, $f_1$ = 4 mm, $f_{cyl1}$ = -20 mm, $f_{cyl2}$ = 50 mm, $f_2$ = 100 mm, $f_3$ = 300 mm, $f_4$ = 75 mm, $M_1$ and $M_2$ are two meniscus mirrors with a radius of curvature of 50 mm.

Figure 2 : Intensity profile of the pump beam measured after the first collimation lens $L_1$. The horizontal direction is the slow axis direction and the vertical direction is the fast axis direction.

Figure 3 : Output power at 607 nm versus incident pump power for 3 different output coupler transmissions.

Figure 4 : Evolution of the beam radius after focalization by a 100 mm lens. Black points : experimental data, black line : numerical fit leading to a $M^2$ value of 1.2. Inset : Intensity beam profile measured at the output of the laser cavity with a CCD camera.

**Figures :**

Figure 1 : Experimental set-up. DL : diode laser, $f_1$ = 4 mm, $f_{cyl1}$ = -20 mm, $f_{cyl2}$ = 50 mm, $f_2$ = 100 mm, $f_3$ = 300 mm, $f_4$ = 75 mm, $M_1$ and $M_2$ are two meniscus mirrors with a radius of curvature of 50 mm.

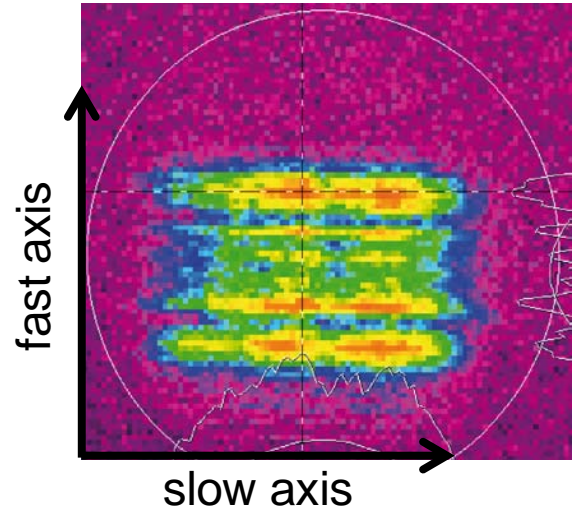

Figure 2 : Intensity profile of the pump beam measured after the first collimation lens L$_1$. The horizontal direction is the slow axis direction and the vertical direction is the fast axis direction.

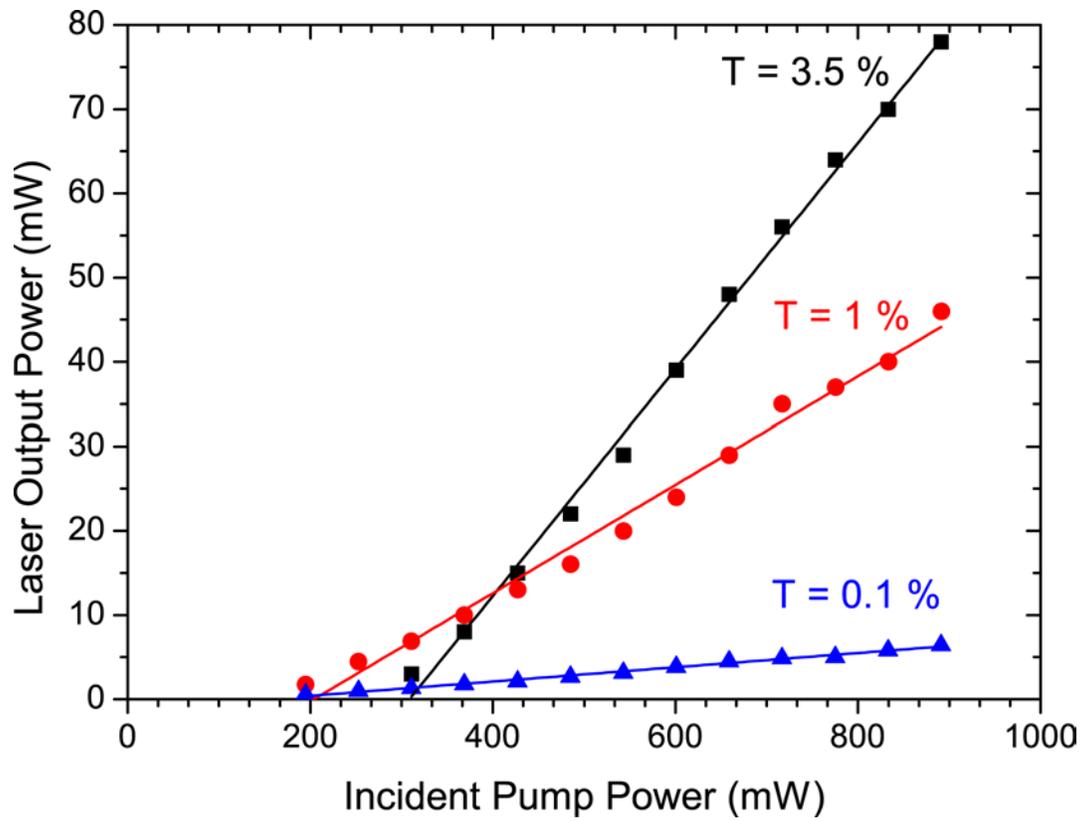

Figure 3 : Output power at 607 nm versus incident pump power for three different output coupler transmissions.

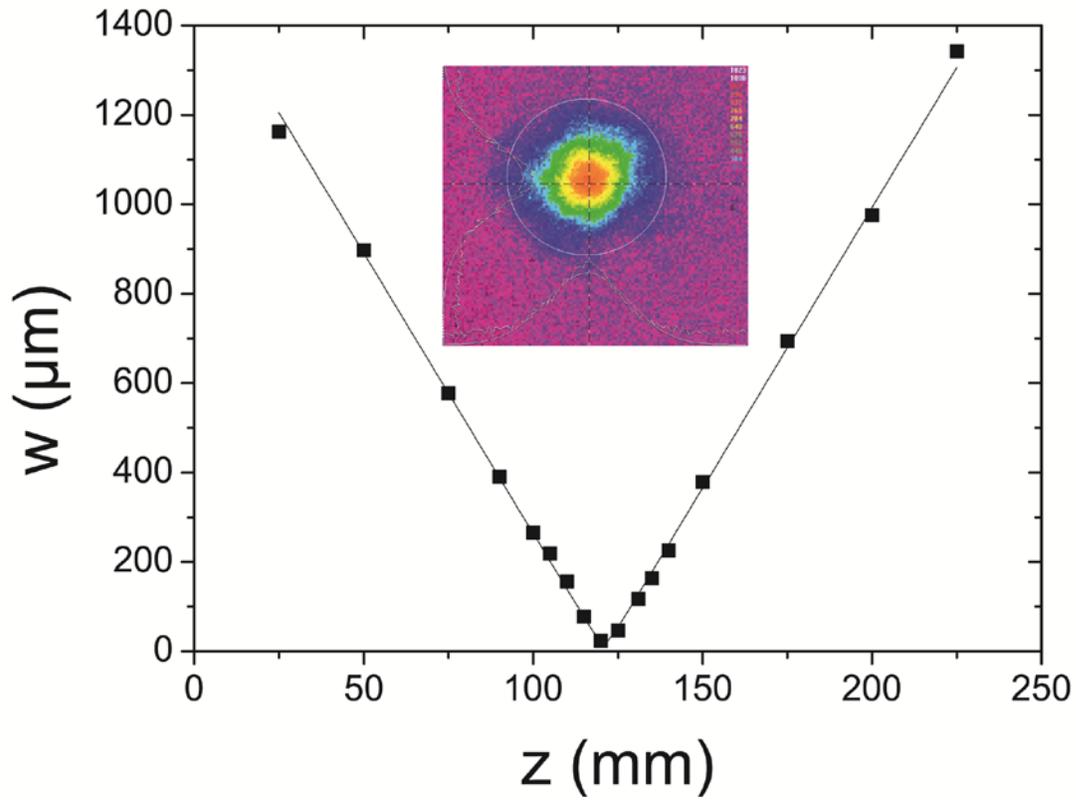

Figure 4 : Evolution of the beam radius after focusing by a 100 mm lens. Black points : experimental data, black line : numerical fit leading to a $M^2$ value of 1.2. Inset : Intensity beam profile measured at the output of the laser cavity with a CCD camera.